\documentclass[preprint,aps]{revtex4}
\usepackage{amssymb}

\usepackage{graphicx}

\begin{document}
\title{Non-local thin films in Casimir force calculations}
\author{R. Esquivel}
\email[]{raul@fisica.unam.mx} \affiliation{Instituto de Fisica, REGINA,
Universidad Nacional Aut\'{o}noma de M\'{e}xico, Apartado Postal
20-364, DF 01000 M\'{e}xico, Mexico}
\author{V. B. Svetovoy}
\email[]{V.B.Svetovoy@el.utwente.nl}
\thanks{On leave from Yaroslavl University, Yaroslavl, Russia}
\affiliation{MESA+ Research Institute, University of Twente, P.O.
217, 7500 AE Enschede, The Netherlands}
\date{\today}

\begin{abstract}
The Casimir force is calculated between plates with thin metallic
coating. Thin films are described with spatially dispersive
(nonlocal) dielectric functions. For thin films the nonlocal effects
are more relevant than for half-spaces. However, it is shown that
even for film thickness smaller than the mean free path for
electrons, the difference between local and nonlocal calculations of
the Casimir force is of the order of a few tenths of a percent. Thus
the local description of thin metallic films is adequate within the
current experimental precision and range of separations.
\end{abstract}

\maketitle

\section{Introduction}

The Casimir force between uncharged metallic plates \cite{Cas48}
(see also reviews \cite{Mil94,Mos97,Kar99,Mil01,Bor01,Mil04})
attracted considerable attention in the past years. The force was
measured in a number of experiments with a high precision using
different techniques and geometric configurations
\cite{Lam97,Moh98,Roy99,Har00,Ede00,Cha01,Bres02,Dec03a,Dec03b}. On
the other hand, the potential applications of the force in micro and
nanomechanics is still largely unexplored. Actuation and nonlinear
behavior of a mechanical oscillator with the Casimir force were
demonstrated \cite{Cha01} and the  importance of the force in
adhesion and stiction has also been discussed
\cite{Buk01,Joh02,Zha03}. Due to technological reasons thin coating
layers or multilayered structures are often in use in
micromechanical devices. The main question to be addressed in this
paper is how important are the nonlocal effects when the film
thickness is smaller than the mean free path of the  electrons.

For the first time the problem of a thin metallic layer on top of
another metal appeared in connection with the first atomic force
microscope (AFM) experiments \cite{Moh98,Roy99}. In these
experiments a relatively thick $Al$ layer was covered with $Au/Pd$
film of 20 nm \cite {Moh98} or 8 nm \cite{Roy99} thick to prevent
aluminum oxidation. Because the film was thin enough to be
transparent for the light with a characteristic frequency $\omega
_{ch}=c/2a$, where $a$ is the distance between the bodies, it was
concluded that $Au/Pd$ layer did not influence on the force\cite{footnote1}. In
actual calculations \cite{Kli99} the thin $Au/Pd$ film was changed
by vacuum. This approach was criticized \cite{Sve00a} on the basis
that according to the Lifshitz formula \cite{Lif56,LP9} the force
depends on the dielectric function $\varepsilon \left( i\zeta
\right) $ at imaginary frequencies $\omega=i\zeta$. Kramers-Kronig
relation shows that at $\zeta _{ch}=c/2a$ low real frequencies
$\omega \ll c/2a$ give significant contribution to
$\varepsilon(i\zeta_{ch}) $. At low frequencies $Au/Pd$ film is not
transparent and it should be taken into account. It was demonstrated
that, indeed, even 8 nm thick film gave significant contribution to
the force. The calculation in Ref. \cite{Kli00} supported this
conclusion but the authors speculated that nonlocal effects due to
small thickness of the film (smaller than the mean free path for
electrons) allowed one to consider the film as transparent.

The question arose again in connection with a recent experiment
\cite{Lis05}, where the force was measured between a plate and
sphere covered with 10 nm or 200 nm $Pd$ film. For thin film the
expected reduction of the force was clearly observed. It was
indicated that spatial dispersion might be important for calculation
of the force in the case of thin film. Also, Bostr\"{o}m and
Sernelius \cite{bostrom2000} pointed out to the need of detailed
studies of nonlocal effects, while studying the retarded van der
Waals force between thin metallic films within a local
approximation.

There have been several works dealing with the problem of
nonlocality in the Casimir force between half spaces. Katz
\cite{katz} was the first to point out the need of a quantitative
study and in his work only a rough estimate of how spatial
dispersion affected dispersive forces was given.  Heindricks
\cite{Heindricks} was able to derive Lifshitz formula in an
approximate way to include nonlocal effects. Similarly, Dubrava
using a phenomenological approach described the Casimir attraction
between  thin films \cite{Dubrava}. More recently, based on the
formalism of nonlocal optics, the effects for thick metallic layers
have been considered. Propagation of bulk plasmons
\cite{Esq03,Esq05,Moch05} and electromagnetic response in the region
of anomalous dispersion \cite{Esq04a} were taken into account,
showing that the spatial dispersion does not contribute
significantly to the Casimir force. The method developed in Ref.
\cite{Esq04a} is very general and can be used for the analysis of
all nonlocal effects including those arising in thin films.

\section{Formalism \label{sec2}}

The Casimir force between two plates separated by a vacuum gap $a$
at a temperature $T$ \ is given by the Lifshitz formula
\cite{Lif56,LP9}. The force is expressed via the reflection
coefficients $R_{1}$ and $R_{2}$ of the plate 1 and 2, respectively,
in the following way:

\begin{equation}
F_{pp}\left( a\right) =-\frac{k_{B}T}{\pi }\sum_{n=0}^{\infty
}{}^{\prime }\int\limits_{0}^{\infty }dqqk_{0}\left[ \left(
R_{1s}^{-1}R_{2s}^{-1}\exp \left( 2ak_{0}\right) -1\right)
^{-1}+\left( R_{1p}^{-1}R_{2p}^{-1}\exp \left( 2ak_{0}\right)
-1\right) ^{-1}\right] ,  \label{Fpp}
\end{equation}

\noindent where subscripts $s$ and $p$ denote the polarization states, ${\bf %
q}$ is the wave vector along the plates, $q=\left| {\bf q}\right| $, and $%
k_{0}$ is the normal component of the wave vector defined as

\begin{equation}
k_{0}=\sqrt{\zeta _{n}^{2}/c^{2}+q^{2}}.  \label{k0}
\end{equation}

\noindent In Eq. (\ref{Fpp}) the sum is calculated over the
Matsubara frequencies

\begin{equation}
\zeta _{n}=\frac{2\pi k_{B}T}{\hbar }n.  \label{Mats}
\end{equation}

\noindent The reflection coefficients $R_{1}$ and $R_{2}$ are different for $%
s$ and $p$ polarizations and are functions of $q$ and imaginary frequencies $%
\zeta _{n}$. They comprise material properties of the plates and
for this reason we start our analysis from the reflection
coefficients.

\subsection{Local case}

To set our notation, we first study briefly the known  local case,
when the optical response depends only on  frequency. We start the
analysis from a thin film of thickness $h$ on a substrate. It will
be assumed here that the film is continuous.

For a film on an infinitely thick substrate the problem is
rather simple. The Maxwell equations are solved with the boundary
conditions which are the continuity of the tangential components of
electric and magnetic fields on both boundaries of the film. The
problem can be solved at real frequencies and then analytically
continued to the imaginary axis. In the local limit the film and
substrate are described by their local dielectric functions
which will be denoted as $\varepsilon _{1}\left( \omega \right) $ and $
\varepsilon _{2}\left( \omega \right) $. In general, the indexes
marking the layers will increase from top to bottom of the plate.
The dielectric function of vacuum will be taken as $\varepsilon
_{0}\left( \omega \right) =1 $. The reflection coefficients in our
case are well known in optics. At imaginary frequencies they are  \cite{Zho95}:

\begin{equation}
R=\frac{r_{01}-r_{21}\exp \left( -2k_{1}h\right)
}{1-r_{01}r_{21}\exp \left( -2k_{1}h\right) },  \label{Refl2}
\end{equation}

\noindent where $r_{ml}$ are the reflection coefficients from the
boundary between media $l$ and $m$. These coefficients depend on
the polarization, $s$ or $p$, and are defined as

\begin{equation}
r_{ml}^{s}=\frac{k_{m}-k_{l}}{k_{m}+k_{l}},\quad r_{ml}^{p}=\frac{%
\varepsilon _{l}k_{m}-\varepsilon _{m}k_{l}}{\varepsilon
_{l}k_{m}+\varepsilon _{m}k_{l}},  \label{reflb}
\end{equation}

\noindent where $k_{m}$ is the normal component of the wave vector
in the medium $m$:

\begin{equation}
k_{m}=\sqrt{\varepsilon _{m}\left( i\zeta \right) \frac{\zeta ^{2}}{c^{2}}%
+q^{2}}.  \label{km}
\end{equation}

\noindent It is easy to check that for $h\rightarrow \infty $ (thick film) $%
R\rightarrow r_{01}$ and in the opposite limit $h\rightarrow 0$
the reflection coefficient coincides with that for the substrate:
$R\rightarrow r_{02}$.

To understand the variation of the Casimir force with the film
thickness, we first study  the behavior of the reflection
coefficients. For a qualitative analysis it will be assumed that a
metal, film or substrate, can be described with the Drude
dielectric function

\begin{equation}
\varepsilon \left( i\zeta \right) =1+\frac{\omega _{p}^{2}}{\zeta
\left( \zeta +\omega _{\tau }\right) },  \label{Drude}
\end{equation}

\noindent where $\omega _{p}$ and $\omega _{\tau }$ are the Drude
parameters which are different for each layer. For thin films,
$\omega_{\tau}$ is a function of the film thickness. This dependence
appears because, in addition to the internal scattering processes, for
thin films scattering from the surfaces is important. These
processes are independent of each other and the relaxation time in
the Drude model is
$\omega_{\tau}=\omega^{bulk}_{\tau}+\omega^{surf}_{\tau}(h) $.  This
effect becomes important when the thickness is smaller than the mean
free path for electron. Dependence on $h$ of $\omega^{surf}_{\tau}$
is explained by the Fuchs-Sondheimer theory \cite{Fuc38, Son52}. When
$h$ is much smaller than the mean free path, this dependence is
given by

\begin{equation}\label{Fuchs}
    \omega^{surf}_{\tau}(h)=\frac{3}{8}(1-p)\frac{v_F}{h},
\end{equation}

\noindent where $v_F$ is the Fermi velocity and an electron has
probability $p$ of being specularly reflected from the surface. As
one can see from Eq. (\ref{Fuchs}) only diffusely reflected
electrons contribute to $\omega^{surf}_{\tau}(h)$. Experimental
results concerning the specularity are far from unique. Very
different values of $p$ in the range $0<p<1$ were used to explain
the experimental results \cite{Fis80}. In this paper we investigate
the nonlocal effects for specular reflection of electrons on the
surface and do not include in the consideration $h$-dependence of
the relaxation frequency. But in any case our results are not very
sensitive to the exact value of $\omega_{\tau}$.

Consider first the system consisting of $SiO_{2}$ substrate with $%
\varepsilon _{2}=4$ and $Au$ film on top of it with the parameters
$\omega _{p}=9.0\ eV$, $\omega _{\tau }=0.035\ eV$ \cite{Lam00}. It
is convenient to introduce dimensionless variables and parameters as
follows:

\begin{equation}  \label{dimless}
\Omega =\frac \zeta {\omega _p},\ Q=\frac{cq}{\omega _p},\ \gamma =\frac{%
\omega _\tau }{\omega _p},\ H=\frac{\omega _ph}c.
\end{equation}

\begin{figure}[tbp]
\includegraphics[width=8.6cm]{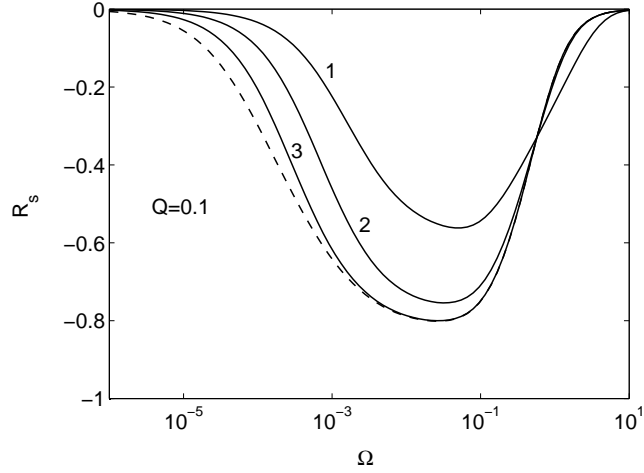}\newline
\caption{Reflection coefficient for $s$-polarization as a function
of $\Omega$ in the local case (metal film on the dielectric
substrate). All the results are presented for $Q=0.1$. Infinitely
thick film is given by the dashed line. The curves 1, 2, and 3
correspond to the dimensionless thickness $H=0.3$, $1$, and $3$,
respectively. }\label{fig1}
\end{figure}

\noindent The reflection coefficient for $s$-polarization as a
function of the dimensionless frequency $\Omega $ is shown in Fig.
\ref{fig1}. The dashed curve corresponds to semi-infinite metal
$h\rightarrow \infty $. It was calculated with $Q=0.1$. This value
is taken for the characteristic wave number $q\sim 1/2a$ at $a\sim
100\ nm$. The solid lines marked as 1, 2, and 3 correspond to the
dimensionless thickness $H=0.3$, $1$, and $3$, respectively. Note
that $H=1$ gives the film thickness $h$ equal to the penetration
depth $\delta =c/\omega _{p}\approx 22\ nm$ ($Au$). One can see that
$R_{s}$ decreases fast with the thickness. When $Q$ increases the
film also becomes more transparent for $s$-polarization. The other
distinctive feature is that $R_s$ is going to zero in the limit
$\Omega\rightarrow0$. In this limit $s$-polarized field degenerates
to pure magnetic field, which penetrate freely via the metallic
film.

\begin{figure}[tbp]
\includegraphics[width=8.6cm]{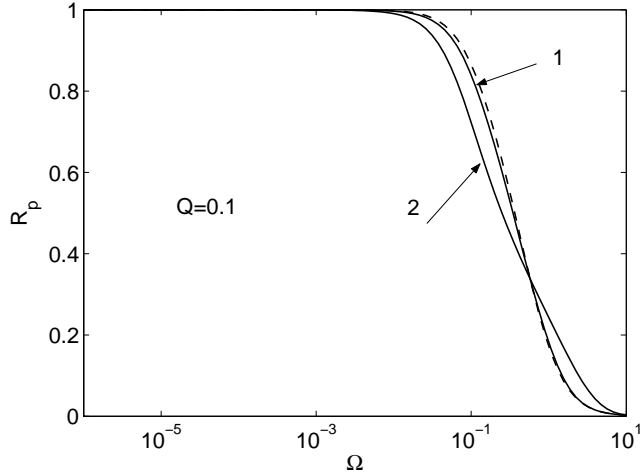}\newline
\caption{Reflection coefficient for $p$-polarization as a function
of $\Omega$ in the local case (metal film on the dielectric
substrate). Infinitely thick layer is shown by the dashed line. The
solid lines marked as 1 and 2 correspond to $H=1$ and $0.1$,
respectively.}\label{fig2}
\end{figure}

The reflection coefficient for $p$-polarization shows a different
behavior as one can see in Fig. \ref{fig2}. The dashed line
represents the thick film and the solid lines marked as 1 and 2
correspond to $H=1$ and $0.1$, respectively. Variation of $R_{p}$
with the film thickness is not very significant. The reason for this
is the effective screening of the  $E_z$ component even by a very
thin metallic layer. An important conclusion can be drawn from this
simple fact. The film thickness affects mostly the contribution of
$s$-polarization, but the part of the force connected with
$p$-polarization is changed weakly in the local case.

\begin{figure}[tbp]
\includegraphics[width=8.6cm]{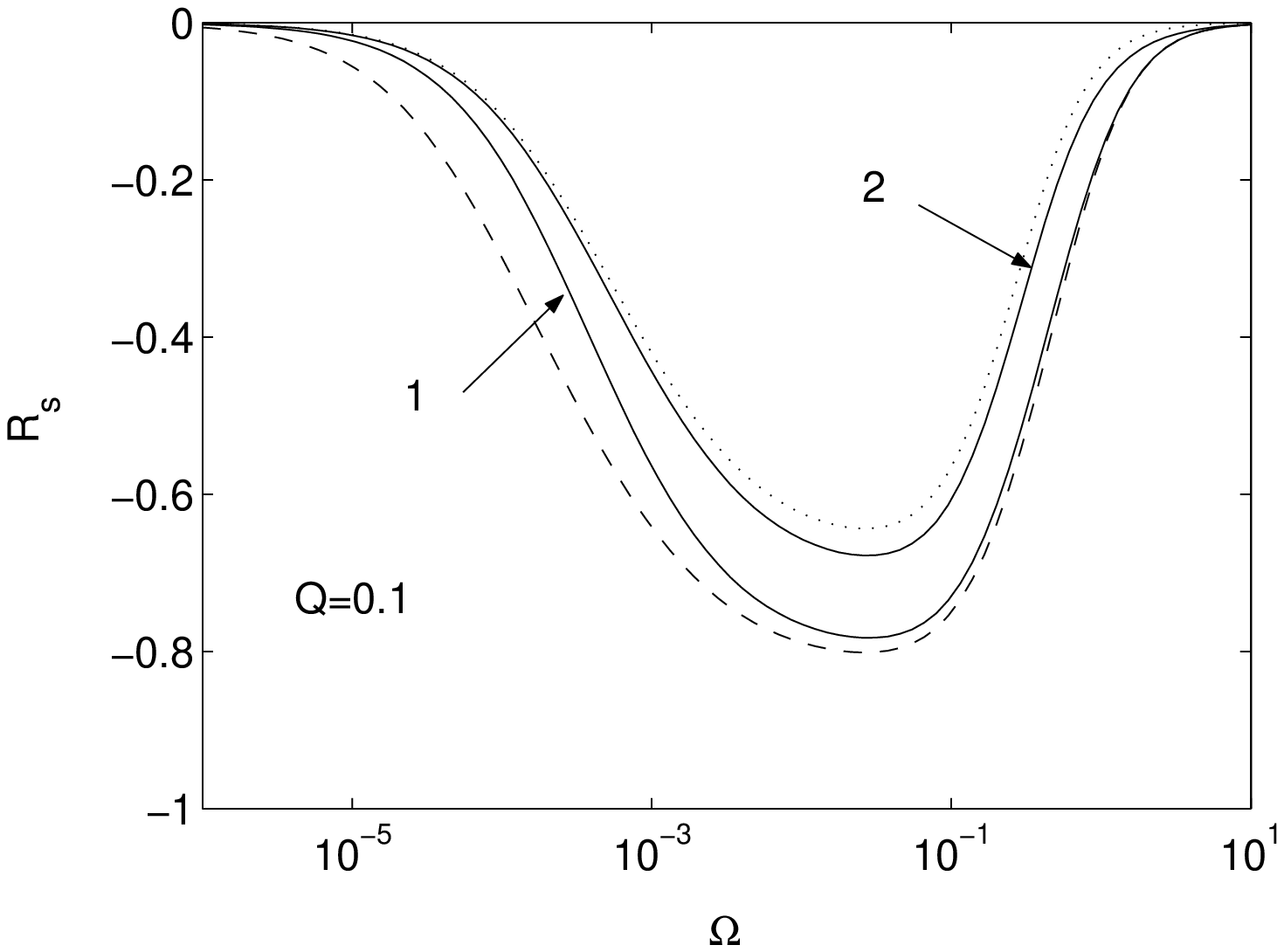}\newline
\caption{Reflection coefficient for $s$-polarization as a function
of $\Omega$ in the local case (metal film on the metallic
substrate). The dotted line represents the substrate, $H=0$, the
dashed line represents the thick top layer, $H\rightarrow\infty$.
The curves 1 and 2 correspond to $H=1$ and $H=0.1$,
respectively.}\label{fig3}
\end{figure}

\begin{figure}[tbp]
\includegraphics[width=8.6cm]{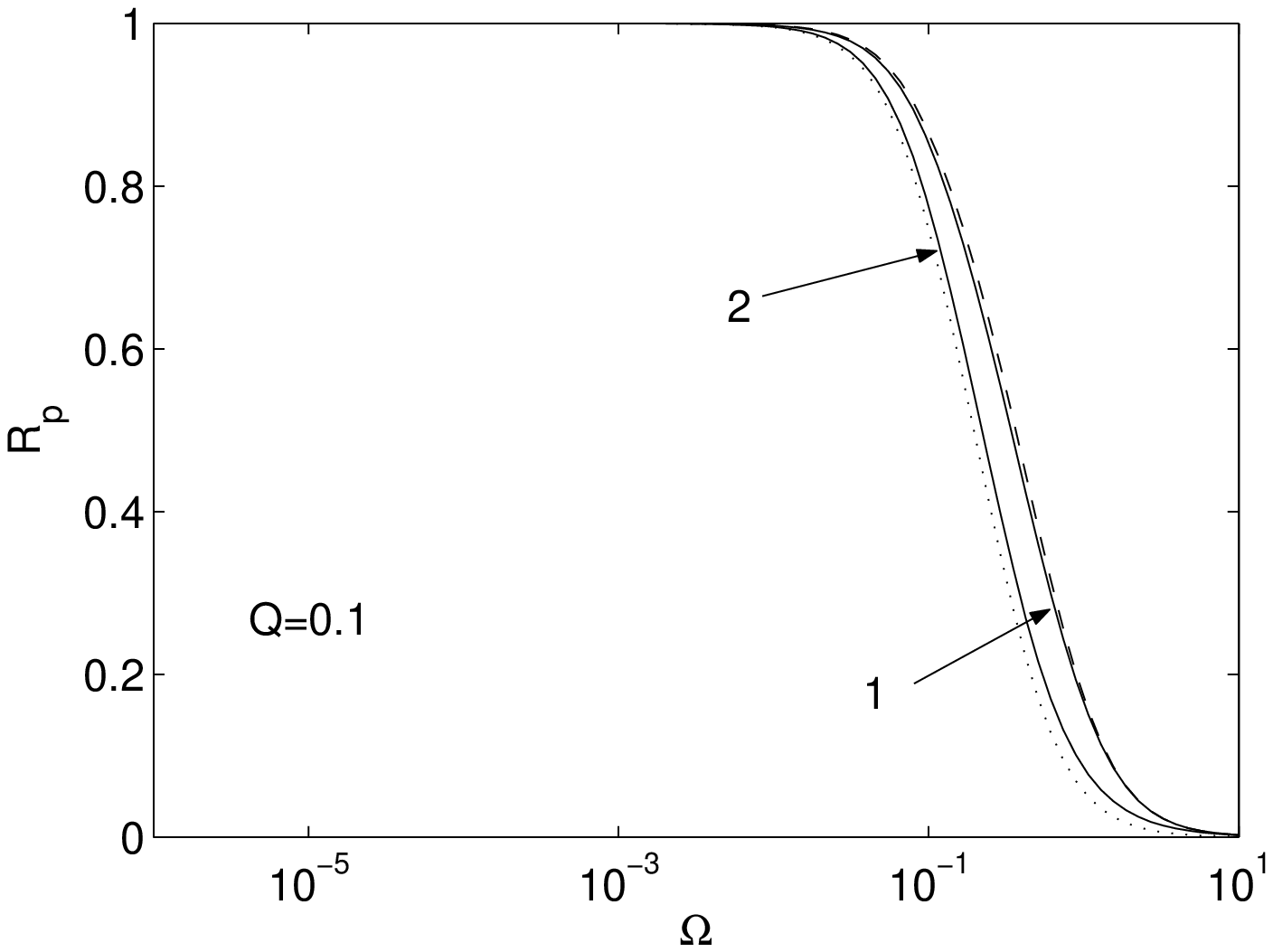}\newline
\caption{The same as Fig. \ref{fig3} but for
$p$-polarization}\label{fig4}
\end{figure}

Consider now the effect of a thin film on top of a thick metallic
layer. It will be assumed that both metals can be described by the
Drude dielectric functions $\varepsilon _{1}\left( i\zeta \right) $
and $\varepsilon _{2}\left( i\zeta \right) $ which differ from each
other only by the values of parameters $\omega _{ip}$ and $\omega
_{i\tau }$ ($i=1,2$). It is clear that in dependence on the film
thickness the reflection coefficients will be in between the lines
describing metal 1 ($h\rightarrow \infty $) or metal 2
($h\rightarrow 0$). In Fig. \ref{fig3} we present the case when the
top layer is better reflector than the bottom one. The dotted line
gives $r_{02}$ and the dashed
line represents $r_{01}$. The results for the film with thickness $H=1$ and $%
0.1$ are marked as 1 and 2, respectively. In our  calculations,
 the ratios $\omega _{1p}/\omega _{2p}=2$ and $\omega
_{1\tau }/\omega _{2\tau }=1$ where used and dimensionless
parameters (\ref{dimless}) were defined relative  to the parameters
of the top layer 1. The relaxation frequencies, $\omega _{i\tau }$,
influence mostly on low frequency behavior of $R_{s}$. They are not
very important for the Casimir force because the main contribution
in the force comes from the imaginary frequencies $\Omega \sim
c/2a\omega _{p}\gg \gamma $ where $\omega _{\tau }$ does not play
significant role. The reflection coefficient, $R_{p}$, for
$p$-polarization is shown in Fig. \ref{fig4}. The curves 1 and 2
correspond $H=1$ and $H=0.1$, respectively. Again one can conclude
that the top layer is more important for $s$ than for
$p$-polarization.

\subsection{Nonlocal case}

For propagating photons the reflectivity of thin films in the
nonlocal case has been analyzed in Ref. \cite{Jon69}. It was assumed
that electrons are reflected specularly on both boundaries of the
film. Let us consider first $s$-polarization. Similar to the case of
a semi-infinite metal \cite {Kli68} the tangential component of the
electric field is considered as even on each boundary:

\begin{equation}  \label{bcs1}
E_y\left( mh-z\right) =+E_y\left( mh+z\right) ,
\end{equation}

\noindent where $z$ is the direction normal to the film surface, $m$
is an arbitrary integer, and the plane of incidence was chosen to be
$x-z$. The Maxwell equations and Eq. (\ref{bcs1}) demand for the
magnetic field on the boundaries the following conditions:

\begin{equation}  \label{bcs2}
H_x\left( mh-z\right) =-H_x\left( mh+z\right) ,\quad H_z\left(
mh-z\right) =+H_z\left( mh+z\right) .
\end{equation}

\noindent Formally the conditions (\ref{bcs1}), (\ref{bcs2})
continue the film of finite thickness to the infinite layer. These
conditions mean that the fields can be considered as periodic with
period $2h$, and they  can be expanded in a Fourier series.

In the nonlocal case the material is characterized by the impedance
instead of local dielectric function. The impedance is defined as
the ratio of tangential components of electric and magnetic fields
just below the surface. For $s$ and $p$-polarizations the impedances
of metallic film were found in Ref. \cite{Jon69} with the method
which is direct generalization of the method used for semi-infinite
layer \cite{Kli68}. The film has two surfaces and the impedances one
can define on each of them:

\begin{equation}
Z_{s}=-\left. \frac{E_{y}}{H_{x}}\right| _{z=\delta ,h-\delta },\
Z_{p}=\left. \frac{E_{x}}{H_{y}}\right| _{z=\delta ,h-\delta },
\label{impdef}
\end{equation}

\noindent where $\delta \rightarrow 0$. It was noted \cite{Jon69}
that instead of impedances (\ref{impdef}) one can use a different
couple for each polarization which can be easy calculated. These new
impedances were introduced as the ratio of the fields even or odd
relative to the film center $z=h/2$. Even or odd fields will be
marked by the superscripts $(1)$ or $(2)$, respectively. The new
impedances

\begin{equation}
Z_{s}^{(1,2)}=-\left. \frac{E_{y}^{(1,2)}}{H_{x}^{(1,2)}}\right|
_{z=\delta },\ Z_{p}^{(1,2)}=\left.
\frac{E_{x}^{(1,2)}}{H_{y}^{(1,2)}}\right| _{z=\delta },
\label{newdef}
\end{equation}

\noindent are the same on both boundaries of the film because of
the symmetry conditions

\begin{equation}
E_{x,y}^{(1)}\left( \delta \right) =E_{x,y}^{(1)}\left( h-\delta
\right) ,\ E_{x,y}^{(2)}\left( \delta \right)
=-E_{x,y}^{(2)}\left( h-\delta \right) \label{symcon}
\end{equation}

\noindent and similarly  for the magnetic field.

Explicit expressions for these impedances were found in Ref.
\cite{Jon69}:

\begin{equation}
Z_{s}^{(1,2)}=i\frac{2\omega }{ch}\sum_{n=(odd,\
even)}\frac{1}{\frac{\omega
^{2}}{c^{2}}\varepsilon _{t}\left( \omega ,k\right) -\left( \frac{n\pi }{h}%
\right) ^{2}-q^{2}},  \label{Zs}
\end{equation}

\begin{equation}
Z_{p}^{(1,2)}=i\frac{2\omega }{ch}\sum_{n=(odd,\
even)}\frac{1}{k^{2}}\left[ \frac{q^{2}}{\frac{\omega
^{2}}{c^{2}}\varepsilon _{l}\left( \omega
,k\right) }+\frac{\left( \frac{n\pi }{h}\right) ^{2}}{\frac{\omega ^{2}}{%
c^{2}}\varepsilon _{t}\left( \omega ,k\right) -\left( \frac{n\pi
}{h}\right) ^{2}-q^{2}}\right] .  \label{Zp}
\end{equation}

\noindent where for even, $(1)$, or odd, $(2)$, fields the sum has
to be calculated over $n=2m+1$ or $n=2m$, respectively. The
transverse dielectric function $\varepsilon _{t}\left( \omega
,k\right) $ contributes to $Z_{s}$. It describes the response of the
material on the electric field transverse to the wave vector ${\bf
k}$. In case of the  $p$-polarization $z$-component of electric
field creates a nonzero charge density in the metal producing the
longitudinal field inside of metal. That is why $Z_{p}$ depends also
on the longitudinal dielectric function $\varepsilon _{l}\left(
\omega ,k\right) $. In general, these functions are nonlocal, so
they depend on both $\omega $ and $k$. The absolute value of the
wave vector ${\bf k}$ \ in Eqs. (\ref{Zs}), (\ref{Zp}) is

\begin{equation}
k=\sqrt{\left( \frac{n\pi }{h}\right) ^{2}+q^{2}}.  \label{wv}
\end{equation}

Let us consider now the reflection and transmission coefficients of
the film on a substrate. Note that in Ref. \cite{Jon69} only a free
standing film was considered. To find these coefficients one has to
match the tangential components of the electric and magnetic fields
outside and inside of the film. We assume for simplicity that the
substrate can be described by a local dielectric function or
equivalently by local impedances. This assumption is justified by
the investigation of nonlocal effects at imaginary frequencies for
semi-infinite metals \cite{Esq04a}. It was demonstrated that in
contrast with the real frequencies the nonlocal effect (anomalous
skin effect) brings only minor influence on the reflection
coefficients. Matching the electric field on both sides of the film
for $s$-polarization one gets

\begin{equation}
\begin{array}{c}
E_{y}^{0}\left( 1+R_{s}\right) =E_{y}^{(1)}\left( \delta \right)
+E_{y}^{(2)}\left( \delta \right) ,
\\
E_{y}^{0}t_{s}e^{ik_{2}h}=E_{y}^{(1)}\left( \delta \right)
-E_{y}^{(2)}\left( \delta \right) ,
\end{array}
\label{match1}
\end{equation}

\noindent where $E_{y}^{0}$ is the incident field, $t_{s}$ is the
transmission coefficient and the symmetry conditions (\ref{symcon})
were taken into account. Similar equations are true for the magnetic
field:

\begin{equation}
\begin{array}{c}
H_{x}^{0}\left( 1-R_{s}\right) =H_{x}^{(1)}\left( \delta \right)
+H_{x}^{(2)}\left( \delta \right) ,
\\
H_{x}^{0}t_{s}\frac{k_{2}}{k_{0}}e^{ik_{2}h}=-H_{x}^{(1)}\left(
\delta \right) +H_{x}^{(2)}\left( \delta \right) .
\end{array}
\label{match2}
\end{equation}

\noindent Eqs. (\ref{match1}) and (\ref{match2}) can be solved for
$R_{s}$ and $t_{s}$ using the impedance definition (\ref{newdef}).
As the result the reflection coefficient can be presented in the
form:

\begin{equation}
R_{s}=\frac{\left( Z_{s1}^{(1)}-Z_{s0}\right) \left(
Z_{s1}^{(2)}+Z_{s2}\right) +\left( Z_{s1}^{(2)}-Z_{s0}\right)
\left( Z_{s1}^{(1)}+Z_{s2}\right) }{\left(
Z_{s1}^{(1)}+Z_{s0}\right) \left( Z_{s1}^{(2)}+Z_{s2}\right)
+\left( Z_{s1}^{(2)}+Z_{s0}\right) \left(
Z_{s1}^{(1)}+Z_{s2}\right) }.  \label{Rs}
\end{equation}

\noindent Here we introduced the following notations :
$Z_{s1}^{(1,2)}$ are the nonlocal impedances of the film given by
Eq. (\ref{newdef}), $Z_{s2}$ is the local impedance of the
substrate defined as

\begin{equation}
Z_{s2}=\frac{\omega }{ck_{2}},  \label{Zs2}
\end{equation}

\noindent and

\begin{equation}
Z_{s0}=\frac{\omega }{ck_{0}},  \label{Zs0}
\end{equation}

\noindent is the ''impedance'' of the plane wave defined as the
ratio of electric and magnetic fields in the wave. The formula
(\ref{Rs}) for $R_{s}$ cannot be presented in the same form
(\ref{Refl2}) as in the local case. This is because we used the
impedances (\ref{newdef}) instead of that given by Eq.
(\ref{impdef}). As we will see both Eqs. (\ref{Refl2}) and
(\ref{Rs}) coincide in the local limit.

In the same way one can find the reflection coefficient for $p$
-polarization, $R_{p}$. In this case the equations similar to
(\ref{match1}), (\ref{match2}) with the interchange
$x\leftrightarrow y$ will be true, the impedance of the plane wave
is defined as

\begin{equation}
Z_{p0}=\frac{ck_{0}}{\omega }=\frac{1}{Z_{s0}},  \label{Zp0}
\end{equation}

\noindent and the local impedance of the substrate is

\begin{equation}
Z_{p2}=\frac{ck_{2}}{\omega \varepsilon _{2}\left( \omega \right)
}. \label{Zp2}
\end{equation}

\noindent The final expression for $R_{p}$ is

\begin{equation}
R_{p}=-\frac{\left( Z_{p1}^{(1)}-Z_{p0}\right) \left(
Z_{p1}^{(2)}+Z_{p2}\right) +\left( Z_{p1}^{(2)}-Z_{p0}\right)
\left( Z_{p1}^{(1)}+Z_{p2}\right) }{\left(
Z_{p1}^{(1)}+Z_{p0}\right) \left( Z_{p1}^{(2)}+Z_{p2}\right)
+\left( Z_{p1}^{(2)}+Z_{p0}\right) \left(
Z_{p1}^{(1)}+Z_{p2}\right) }.  \label{Rp}
\end{equation}

\noindent It differs from Eq. (\ref{Rs}) only by the general sign
and the change $s\rightarrow p$.

If the substrate is changed by vacuum, $Z_{\alpha 2}\rightarrow
Z_{\alpha 0}$ ($\alpha =s,\ p$), we reproduce the reflection
coefficient found in Ref. \cite{Jon69}:

\begin{equation}
R_{\alpha }=\frac{1}{2}\left( r_{\alpha }^{(1)}+r_{\alpha
}^{(2)}\right) ,\quad \alpha =s,\ p,  \label{rts}
\end{equation}

\noindent where the ''partial'' reflection coefficients are
connected with the impedances by the usual relations

\begin{equation}
r_{s}^{\left( 1,2\right) }=-\frac{Z_{s0}-Z_{s}^{\left( 1,2\right) }}{%
Z_{s0}+Z_{s}^{\left( 1,2\right) }},\quad r_{p}^{\left( 1,2\right) }=\frac{%
Z_{p0}-Z_{p}^{\left( 1,2\right) }}{Z_{p0}+Z_{p}^{\left( 1,2\right)
}}. \label{rspart}
\end{equation}

\noindent In the local limit both the transverse $\varepsilon
_{t}$ and longitudinal $\varepsilon _{l}$ dielectric functions
coincide with the local function: $\varepsilon _{t}\left( \omega
,k\right) \rightarrow \varepsilon _{l}\left( \omega ,k\right)
\rightarrow \varepsilon _{1}\left( \omega \right) $. In this case
the sums in Eqs. (\ref{Zs}), (\ref{Zp}) can be found explicitly.
For example, for $s$-polarization one has

\begin{equation}
Z_{1s}^{(1),\ loc}=-i\frac{\omega }{ck_{1}}\tan
\frac{hk_{1}}{2},\quad Z_{1s}^{(2),\ loc}=i\frac{\omega
}{ck_{1}}\cot \frac{hk_{1}}{2}. \label{Zsloc}
\end{equation}

\noindent Substituting it in Eq. (\ref{Rs}) one can check that the
reflection coefficient for the local case given by Eq. (\ref{Refl2})
is reproduced.

All the equations above were written for real frequencies.
Transition to imaginary frequencies, which are the main point of our
interest, can be done by a simple analytic continuation. To get the
nonlocal effects in the reflection coefficients, we have to fix  the
nonlocal dielectric functions. At imaginary frequencies in the
Boltzmann approximation they are given by the relations
\cite{Esq04a}

\begin{equation}  \label{El}
\varepsilon _l\left( \Omega ,v\right) =1+\frac{f_l\left( v\right)
}{\Omega \left( \Omega +\gamma \right) },\quad f_l\left( v\right)
=\frac 3{v^2}\cdot \frac{v-\arctan v}{v+\frac \gamma \Omega \left(
v-\arctan v\right) },
\end{equation}

\begin{equation}  \label{Et}
\varepsilon _t\left( \Omega ,v\right) =1+\frac{f_t\left( v\right)
}{\Omega \left( \Omega +\gamma \right) },\quad f_t\left( v\right)
=\frac 3{2v^3}\left[ -v+\left( 1+v^2\right) \arctan v\right] ,
\end{equation}

\begin{equation}  \label{vdef}
v=\frac{v_F}c\frac{\sqrt{\left( \frac{n\pi }H\right) ^2+Q^2}}{\Omega
+\gamma },
\end{equation}

\noindent where $v_{F}$ is the Fermi velocity. The dimensionless
variables (\ref{dimless}) have been introduced in Eqs.
(\ref{El})-(\ref{vdef}) . In addition, we have neglected in Eqs.
(\ref{El}) -(\ref{Et})  the contribution due to the interband
transitions.

\begin{figure}[tbp]
\includegraphics[width=8.6cm]{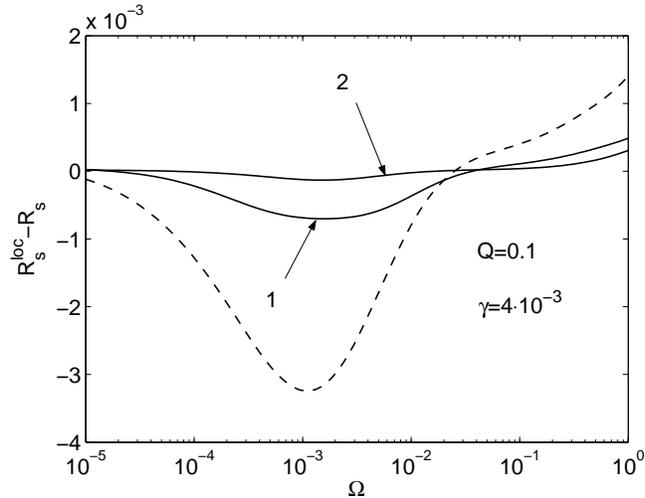}\newline
\caption{Difference between the reflection coefficients for
$s$-polarization in the local and nonlocal cases (metallic film on
the dielectric substrate). The dashed curve was calculated for
very thick film. The solid curves are presented for $H=1$ (1) and
$H=0.1$ (2). }\label{fig5}
\end{figure}

The reflection coefficients in the nonlocal case were calculated
numerically. In Fig. \ref{fig5} the difference between local and
nonlocal coefficients $R_{s}$ is shown for $Au$ film on top of
$SiO_2$ substrate. The dashed curve corresponds to very thick film,
$H\rightarrow \infty $. The solid lines marked as 1 and 2 are
presented for $ H=1$ and 0.1, respectively.  As before, $H=1$
corresponds to the penetration depth of $Au$ ($\delta=22 $ nm). The
thick film clearly demonstrates the anomalous skin effect at $\Omega
\sim \gamma $, although the magnitude of the effect is small as was
already noted in Ref. \cite{Esq04a}. Even this small effect
decreases with the film thickness as the curves 1 and 2 show. The
nonlocal effect increases with $Q$ but it is smaller than 1\% even
for $Q=1$. It should be noted that the Boltzmann approximation is
good while $ \Omega <1$, but when $\Omega $ approaching 1 the
reflection coefficient itself becomes small and there is no sense to
keep the nonlocal correction in this range. Similar result was found
for the film on top of a metallic substrate. One can conclude that
for $s$-polarization the nonlocal effect in the reflection
coefficient is very small and can be neglected in calculation of the
Casimir force.

\begin{figure}[tbp]
\includegraphics[width=8.6cm]{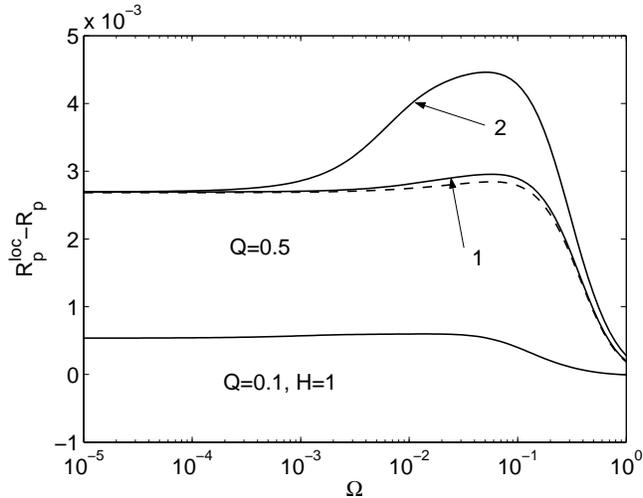}\newline
\caption{Difference between the reflection coefficients for
$p$-polarization in the local and nonlocal cases (metal film on
the metallic substrate). The lower line is given for $Q=0.1, H=1$.
The upper series is for $Q=0.5$. The dashed curve presents very
thick film. The solid curves are given for $H=1$ (1) and $H=0.1$
(2). }\label{fig6}
\end{figure}

The situation for $p$-polarization is shown in Fig. \ref{fig6} for
the film on top of metallic substrate. As in the local case the
substrate was chosen to have the plasma frequency 2 times smaller
than that for the film. The lower curve corresponds to $Q=0.1$ and
$H=1$. The upper series of curves is given for $Q=0.5$.  As
one can see, the nonlocal effect manifests itself in a wider
frequency range and does not disappear even for zero frequency. The
latter is the result of Thomas-Fermi screening as was explained in
Ref. \cite{Esq04a}. The effect is still small but the nonlocal
contribution in the Casimir force will be larger than that for
$s$-polarization. This is because the nonlocal effect is the largest
at frequencies which give the main contribution in the Casimir
force.

\section{ Effects of Spatial Dispersion on the Casimir force }

To quantify the effect of spatial dispersion on the Casimir force,
we calculate the percent difference between the local case and
nonlocal case ($\Delta\%=|(F_{local}-F_{nonlocal})/F_{local}|$), as
a function of separation.

\begin{figure}[tbp]
\includegraphics[width=8.6cm]{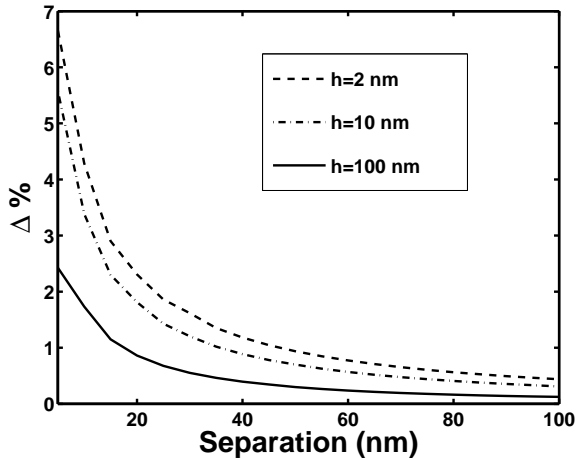}\newline
\caption{Percent difference $\Delta\%$ between the local and
nonlocal Casimir force between  $Au$ free standing films. The values
for the thick film coincide with those obtained for half-spaces
\cite{Esq04a}. }\label{fig7}
\end{figure}

First we consider the case of free standing metallic films. The
system is similar to that considered by Bostr\"{o}m and Sernelius
\cite{bostrom2000}. The percent difference $\Delta\%$ as a function
of separation is presented in Fig. \ref{fig7}, for three different
thicknesses. The results for the thick film $h=100\ nm$ coincides
with the results obtained for half spaces in our previous work
\cite{Esq04a}. As the thickness decreases the nonlocal effects
become more relevant. Thin films have a more complicated nonlocal response than half spaces. For $p$-polarized waves, 
surface plasmons on each side of the film can interfere \cite{lopez82}, creating standing waves that will increase the electromagnetic 
absorption of the field  that will decrease the Casimir force. 
 These resonance conditions are evident from Eq.
(\ref{wv}) where $k_z=n\pi/L$.

\begin{figure}[tbp]
\includegraphics[width=8.6cm]{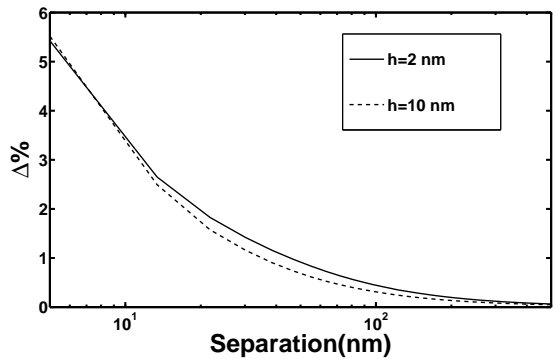}\newline
\caption{Percent difference $\Delta\%$ between the local and
nonlocal Casimir force between two $Au$ films deposited on a
dielectric substrate.}\label{fig8}
\end{figure}

The force is not affected significantly when the thin films are on
substrates. In Figure \ref{fig8} we have plotted the percent
difference between two thin $Au$ films, each deposited on a
dielectric substrate. Again, we assumed $\epsilon=4$ for the
dielectric, just as an illustrative example of the effect of
substrate. The substrates reduce slightly the value of $\Delta \%$
for both curves shown, with the obvious limit that when the
substrate has the same dielectric function as the film, we recover
the results for the force between half-spaces. This means that the
effect of the substrate is to allow energy transfer out of the thin
film into the substrate.

\begin{figure}[tbp]
\includegraphics[width=8.6cm]{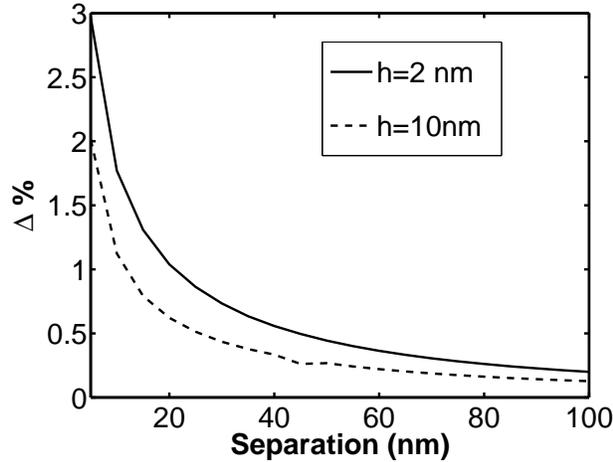}\newline
\caption{Percent difference $\Delta\%$ between the local and
nonlocal Casimir force between $Au$ half space and $Au$ coated
dielectric. }\label{fig9}
\end{figure}

The difference between the local and nonlocal cases can be reduced
in a system consisting of $Au$ half space and $Au$ coated substrate.
Again, we took a dielectric ($\epsilon=4$). The effect of spatial
dispersion reduces significantly as compared to the cases treated in
Figs. \ref{fig7} and \ref{fig8}. This shows that the most important
part of the spatial dispersion effect come from the thin films. If
in current experiments the separation can go down to $50\ nm$, in
the system shown in Fig. \ref{fig9}, the nonlocal correction is of
the order of $0.34\%$.

The result holds for different substrates. This is shown in Table 1,
where we presented the percent difference between the local and
nonlocal forces for $Au$ film  deposited on different substrates.
All data are given for a separation of 50 nm.  As before,
$\omega_{p1}/\omega_{p2}$ is the ratio of the $Au$ plasma frequency
to that of the metallic substrate, assuming the damping factor
remains the same. The case $\varepsilon=1$ corresponds to the free
standing $Au$ thin film (no substrate).

\begin{table}
 \caption{The magnitude of the nonlocal effect for different substrates at
 a fixed separation of 50 nm and a film thickness of 2 nm. }
 \begin{tabular}{|c|c|}\hline
 substrate&$\Delta\%$ \\
 dielectric, $\epsilon=4$&0.34\\
 metal, $\frac{\omega_{p1}}{\omega_{p2}}=2$&0.37\\
 metal, $\frac{\omega_{p1}}{\omega_{p2}}=0.5$&0.44\\
 no substrate, $\epsilon=1$&0.44\\  \hline
 \end{tabular}
 \end{table}

\section{Conclusions}

The role of thin metallic coatings in the calculation of Casimir
forces has been studied taking into account  spatial dispersion. The
description of the nonlocal response of thin films is based on the
Kliewer and Fuchs formalism that imposes a symmetrical behavior of
the fields inside thin films. The study of the reflectivities shows
that the main contribution to the nonlocal effect comes from
$p$-polarized light that excites normal modes within the material.
At very small separations, the effects can be appreciable but at
best a percent difference of $7\%$ is found. However, for typical
experimental setups and separations the percent difference between
the local and nonlocal case is of the order of $0.4\%$, that can be
regarded as negligible within current experimental precisions and
the local description is good enough. The effect of thin films
within a local approximation has been measured recently by Lisanti
et al. \cite{Lis05}.

Along with the previous works on nonlocal effects between
half-spaces \cite{Esq03,Esq04a,Esq05,Moch05}, we can generally
conclude that these effects will be difficult to detect at the
current experimental precision. Our results indicate a decrease in
the force due to spatial dispersion. However for half-spaces within
a jellium model it has been shown \cite{Moch05} that the force can
increase due to nonlocal effects because of decrease in the
separation of the optical surfaces that might not coincide with the
physical surface.

\acknowledgements{Partial support from CONACyT project:  44306 and DGAPA-UNAM IN-101605.  We 
thank W.L. Mochan and C. Villarreal for helpful discusions. }


\begin{references}
\bibitem{footnote1}  For most of the experiments, the force is measured between a large sphere   and a plane, being $a$  the closest distance between the sphere's surface and the plane.  In the rest of this work we will refer to $a$ also as the separation between two parallel planes.
\end{references}
\end{document}